\documentclass[sigplan,screen]{acmart}

\setcopyright{rightsretained}
\acmPrice{}
\acmDOI{10.1145/3372885.3373813}
\acmYear{2020}
\copyrightyear{2020}
\acmISBN{978-1-4503-7097-4/20/01}
\acmConference[CPP '20]{Proceedings of the 9th ACM SIGPLAN International Conference on Certified Programs and Proofs}{January 20--21, 2020}{New Orleans, LA, USA}
\acmBooktitle{Proceedings of the 9th ACM SIGPLAN International Conference on Certified Programs and Proofs (CPP '20), January 20--21, 2020, New Orleans, LA, USA}



\usepackage{listings,lstlangcoq}
\usepackage{xspace}
\usepackage[normalem]{ulem}
\usepackage{mathtools}
\usepackage{proof}
\usepackage[inline]{enumitem}
\usepackage{tikz}
\usepackage{balance}

\usetikzlibrary{arrows, positioning, automata}

\lstdefinestyle{customcoq}{
  mathescape=true,
  belowcaptionskip=1\baselineskip,
  breaklines=true,
  xleftmargin=\parindent,
  language=Coq,
  morekeywords={Variant, fun, Arguments, Type, cofix},
  emph={%
    SOCKAPI,ITree,data_at,data_at_
  },
  emphstyle={\bfseries\color{green!40!red!80}},
  showstringspaces=false,
  basicstyle=\small\ttfamily,
  keywordstyle=\bfseries\color{green!40!black},
  commentstyle=\itshape\color{red!40!black},
  identifierstyle=\color{violet!80!black},
  stringstyle=\color{orange},
}

\newenvironment{centermath}
 {\begin{center}$\displaystyle}
 {$\end{center}}

\makeatletter
\newcommand{\specialcell}[1]{\ifmeasuring@#1\else\omit$\displaystyle#1$\ignorespaces\fi}
\makeatother

\newcommand{\ilc}[1]{\mbox{\lstinline[style=customcoq,columns=fixed,basewidth=.48em]{#1}}}

\newif\ifcomments\commentsfalse   
\ifcomments
\newcommand{\proposecut}[1]{\sout{#1}}
\newcommand{\yz}[1]{\textcolor{blue}{{[YZ:~#1]}}}
\newcommand{\sz}[1]{\textcolor{brown!100!black!100}{{[SZ:~#1]}}}
\newcommand{\ph}[1]{\textcolor{green!50!black!100}{{[PH:~#1]}}}
\newcommand{\gh}[1]{\textcolor{white!50!blue!100}{{[GH:~#1]}}}
\else
\newcommand{\proposecut}[1]{}
\newcommand{\yz}[1]{}
\newcommand{\sz}[1]{}
\newcommand{\ph}[1]{}
\newcommand{\gh}[1]{}
\fi

\newcommand{\sxrightarrow}[2][]{%
  \mathrel{\text{$\xrightarrow[#1]{#2}$}}%
}

\newcommand\iffv[1]{\stackrel{\mathclap{\normalfont\mbox{\tiny{#1}}}}{\iff}}

\newcommand{\mysmall}[1]{{\normalfont\mbox{\small{#1}}}}

\newcommand\impliedbyv[1]{\stackrel{\mathclap{\normalfont\mbox{\tiny{#1}}}}{\impliedby}}

\newcommand*{\defeq}{\stackrel{\text{def}}{=}}
\newcommand{\mon}{{\footnotesize\sxrightarrow{m\hspace*{-1pt}o\hspace*{-1pt}n}}}

\newcommand\cat{\mathbin{+\mkern-8mu+}}

\newcommand{\pacon}{\texttt{paco}\xspace}
\newcommand{\gpacon}{\texttt{gpaco}\xspace}

\newcommand{\gfp}[1]{\nu. #1}
\newcommand{\paco}[2]{G_{#1}~#2}
\newcommand{\gpaconoclo}[3]{\hat G_{#1}~#2~#3}
\newcommand{\gpaco}[4]{\hat G_{#1}^{#2}~#3~#4}

\newcommand{\gupaco}[3]{\bar G_{#1}^{#2}~#3}

\newcommand{\bisim}{\sim}
\newcommand{\eutt}{\approx}
\newcommand{\euttge}{\gtrsim}
\newcommand{\Ret}{\ensuremath{\epsilon}\xspace}
\newcommand{\Tau}{\ensuremath{\tau}\xspace}
\newcommand{\Vis}[1]{\ensuremath{\beta(#1)}\xspace}

\newcommand{\bisimF}{\ensuremath{\mathtt{bisimF}}\xspace}
\newcommand{\euttG}[4]{\ensuremath{\mathtt{euttG}~#1~#2~#3~#4}}
\newcommand{\euttGn}{\ensuremath{\mathtt{euttG}}}
\newcommand{\rw}{r_{\tau}}
\newcommand{\rs}{r_{\beta}}
\newcommand{\gw}{g_{\tau}}
\newcommand{\gs}{g_{\beta}}
\newcommand{\euttGg}{\ensuremath{\euttG{\rs}{\rw}{\gs}{\gw}}}

\newcommand{\transU}{\ensuremath{\mathcal{U}}}
\newcommand{\transD}{\ensuremath{\mathcal{D}}}
\newcommand{\concatC}{\ensuremath{\mathcal{C}}}
\newcommand{\euttVC}[1]{\ensuremath{\mathcal{V}_{#1}}}
\newcommand{\betaclo}{\ensuremath{clo_{\beta}}}
\newcommand{\euttF}{\ensuremath{\mathtt{euttF}}\xspace}
\newcommand{\euttFg}{\ensuremath{\mathtt{euttF}~\betaclo}\xspace}
\newcommand{\cpn}{\mathtt{cpn}}
\newcommand{\gres}{\mathtt{gres}}

\newcommand{\utt}{up-to-tau\xspace}
\newcommand{\bfalse}{\mathtt{false}}
\newcommand{\btrue}{\mathtt{true}}

\newcommand{\bclo}{\ensuremath{\mathit{bclo}}}

\begin{abstract}
  Coinductive reasoning about infinitary structures such as streams is widely
  applicable.  However, practical frameworks for developing coinductive proofs
  and finding reasoning principles that help structure such proofs remain a
  challenge, especially in the context of machine-checked formalization.

  This paper gives a novel presentation of an equational theory for reasoning
  about structures up to weak bisimulation.  The theory is both compositional,
  making it suitable for defining general-purpose lemmas, and also incremental,
  meaning that the bisimulation can be created interactively.
  To prove the theory's soundness, this paper also introduces
  \textit{generalized parameterized coinduction}, which addresses expressivity
  problems of earlier works and provides a practical framework for coinductive
  reasoning.  The paper presents the resulting equational theory for streams,
  but the technique applies to other structures too.

  All of the results in this paper have been proved in Coq, and the generalized
  parameterized coinduction framework is available as a Coq library.
\end{abstract}

\begin{document}

\title{An Equational Theory for Weak Bisimulation via Generalized Parameterized Coinduction}

\author[Y. Zakowski]{Yannick Zakowski}
\affiliation{
  \institution{University of Pennsylvania}
  \city{Philadelphia}\state{PA}
  \country{USA}}
\author[P. He]{Paul He}
\affiliation{
  \institution{University of Pennsylvania}
  \city{Philadelphia}\state{PA}
  \country{USA}}
\author[C. Hur]{Chung-Kil Hur}
\affiliation{
  \institution{Seoul National University}
  \city{Seoul}
  \country{Republic of Korea}}
\author[S. Zdancewic]{Steve Zdancewic}
\affiliation{
  \institution{University of Pennsylvania}
  \city{Philadelphia}\state{PA}
  \country{USA}}

\maketitle

\section{Introduction}
\label{sec:intro}

Coinduction is a powerful technique for reasoning about streams, computation
trees, and other infinitary structures that are used widely in semantics and
systems modeling.  As such, coinductive proofs play a significant role in Coq
developments like CompCert~\cite{compcert}, FreeSpec~\cite{freespec}, or
Interaction Trees~\cite{itrees}. \sz{What other citations here? Also, mention
  coinduction in other proof developments? Agda? Isabelle/HOL?}

In such contexts, working with \textit{weak bisimulation} (equivalence modulo
hidden ``internal'' computation steps) is often desirable.  However, na\"{i}ve
ways of applying coinduction, including its use for establishing weak
bisimulations, suffer from lack of compositionality or
incrementality. \textit{Compositionality} allows the proof developer to create
modular proofs using generic lemmas, while still ensuring sound coinductive
reasoning.  \textit{Incrementality} lets them construct the bisimulation
relation by accumulating parts of it during the proof, rather than having to
posit the entire relation up front at the proof's outset.  Both of these properties
are particularly useful in the context of mechanized formal proof.

The situation was improved by the introduction of the parameterized coinduction
approach by \citeN{paco}, and its implementation in the \pacon
library for Coq. The crux of the approach is to move away from specifying the
greatest fixed point up front and instead to work with a predicate parameterized
by ``accumulated knowledge'' that one can use during the construction of the
proof to incrementally build the postfixed point. \ph{TODO: standardize our use of spacing and whatnot for fixed point/fixpoint.}
Hur et al. show that \pacon supports reasoning up-to closures too, and they hinted that it might be
pragmatic to systematically work with the \textit{greatest compatible closure}
(that is, the most general closure among a class satisfying good closure
properties). This idea has been studied in greater length by \citeN{Pous16},
leading to the so-called \emph{companion} approach, to which we compare ourselves in Section~\ref{sec:related}.

Despite these advances, there are still several difficulties with developing
coinductive proofs in interactive theorem provers.  Firstly, the \pacon reasoning
principles are still too weak, resulting in cumbersome proofs. The limitation is
particularly apparent when a proof nests two cofixed points: the inner cofixed point
forgets all available accumulated knowledge, leading to redundant reasoning.
Secondly, the support for up-to reasoning remains either ad hoc or difficult to manipulate
in existing approaches: here we advocate for internalizing and manipulating concretely
defined closures, as opposed to the greatest compatible one.
Finally, it still remains to package coinductive reasoning
principles into ``proof patterns'' for weak bisimulation that are expressive and
easy to work with in practice.

This paper addresses the above problems by making two technical contributions:
\begin{itemize}
\item We present an equational theory over streams that gives a novel axiomatic
  interface for working with weak bisimulations. This yields an ``API,''
  realized by a set of lemmas, that helps users structure their coinductive
  proofs of weak bisimulation.\sz{What is a succint way to explain why this API
    is good?}  This equational theory is a simplified (and self-contained)
  presentation of a formalization of the equational theory of interaction
  trees~\cite{itrees}.

\item To prove the soundness of the equational theory, we introduce
  \emph{Generalized Parameterized Coinduction}, \gpacon, a backwards-compatible
  generalization of the \pacon framework.  This new construction provides the
  ability to record previously available knowledge that has been accumulated
  during a coinductive proof, which solves \pacon's issue with nested cofixed
  points.  Additionally, it has intrinsic support for up-to reasoning, which, in
  contrast to the companion approach, allows for the creation of generic lemmas
  that aid in developing modular proof.  We show that \gpacon supports novel
  coinductive principles.

\end{itemize}

The rest of the paper explains these contributions in detail, working from
\gpacon to the equational theory.  We first briefly review \pacon in
Section~\ref{sec:background} and highlight, by way of example, the shortcomings
that motivate our generalized definition.
Section~\ref{sec:gpaco} presents generalized parameterized coinduction, establishes its basic properties, and explains the reasoning principles that it justifies.
We then incorporate ``up-to closures'' into the definition, again establishing the appropriate metatheory.
Sections~\ref{sec:bisim} and \ref{sec:euttG} apply \gpacon to develop an equational theory for reasoning about (weak) bisimulations of streams with $\tau$ (internal) events.
Here we also present our novel proof rules for working with those bisimulations.
We also show the problem with working with the companion when trying to define these rules.
Section~\ref{sec:implementation} details the implementation of our reasoning principles in Coq.
Finally, Section~\ref{sec:related} provides a comparison with related work.

The reasoning principles presented in this paper are applicable with
little-to-no overhead in the Coq proof assistant through an extension of the
\pacon library.
All of the definitions, metatheory and examples presented here have been verified in Coq.
However, none of it is specific to this proof assistant, and all results
should be transferable to any other system providing support for coinduction.

\section{Background: \pacon and a Motivating Example}
\label{sec:background}

\subsection{Notations}

In this and the following sections, we consider a complete lattice $(C,\sqsubseteq, \sqcup)$
and $f \in C\mon C$, a monotone function over $C$ that we refer to as a functor.
The typical use case in our context will instantiate $C$ with $\mathcal P(T \times T)$ for some type $T$
(\textit{i.e.} the lattice of binary relations over $T$), but the theory applies to any such lattice.
In our Coq formalization, the main lattice is the one of propositional relations over $C:$ \ilc{C -> C -> Prop}.

Write $X_f$ for the set of postfixed points of $f,$ i.e. $x$ such that $x \sqsubseteq f(x)$.
Tarski's theorem implies that $X_f$ admits an upper bound. We write $\gfp{f}$ for this upper bound. Additionally,
this upper bound is the greatest fixed point of $f$, i.e. in particular $\gfp{f} = f(\gfp{f}).$

\subsection{Parameterized Coinduction}

We briefly recall the central idea behind parameterized coinduction and its
reasoning principles.  Intuitively, it consists in moving away from using
$\gfp{f}$ itself and instead conducting a proof toward some $G_{f} \in C\mon C$
that is parameterized by some accumulated knowledge:
\begin{definition}[Parameterized greatest fixed point]
  Define $G \in (C\mon C) \mon (C \mon C)$ to be:
  $$\paco{f}{r} \defeq \gfp{(\lambda y.f(r\sqcup y))} $$
\end{definition}

Here, we think of $r$ as the ``knowledge'' accumulated during a proof.
The intuition and usefulness behind this definition is best illustrated by the
equations it satisfies. The soundness of the approach comes from the fact that
it coincides with the greatest fixed point when no knowledge has been accumulated.

\begin{lemma}[\textsc{Init}]
  $\gfp{f} \equiv \paco{f}{\bot}$
\end{lemma}

The central coinduction principle, mapping to a strong variant of Tarski's
principle, is expressed as an unfolding lemma. It intuitively states that the
coinduction hypothesis as well as the accumulated knowledge are accessible
behind the guard, i.e. an iteration of the functor $f$.

\begin{lemma}[\textsc{Unfold}]
  $\paco{f}{r} \equiv f(r \sqcup \paco{f}{r})$
\end{lemma}

Finally, the accumulation principle is the key to allow for incremental
coinductive proofs: one can enrich the currently accumulated knowledge at any
point.

\begin{lemma}[\textsc{Acc}]
	$y \sqsubseteq \paco{f}{r} \iff y \sqsubseteq \paco{f}{(r\sqcup y)}$
\end{lemma}

The technique has been a wild success, most notably in the context of the Coq proof assistant in which it has been implemented.
It at once enabled both incremental and compositional reasoning principles, two improvements that are of particular value when conducting mechanized proofs.
Notably, parameterized coinduction is also entirely compatible with automation, something that the native reasoning principles provided by Coq for coinduction prohibited in practice.

\subsection{Example: \pacon{}'s Shortcomings}

The typical coinductive proof using \pacon aims to prove a goal of the
form $y \sqsubseteq \gfp{f}$. One starts by using \textsc{Init} to obtain
$y \sqsubseteq \paco{f}{\bot}$, after which the proof proceeds by using
\textsc{Unfold} and \textsc{Acc} interleaved with other steps of
equational reasoning. Such incremental proofs are considerably simpler to
construct in an interactive theorem prover. However, the \pacon{} lemmas falter
in the presence of nested cofixed points: they lose too much information about
the accumulated knowledge, leading to redundant and more awkward to construct
proofs, a deficiency that becomes more problematic as the technique scales to
reason about more complex systems.

\newcommand{\app}{}
\newcommand{\Viss}[1]{\ensuremath{#1}}

To illustrate this phenomenon, consider the coinductive stream
(or lazy list, since these streams can also be finite) data type
that might be used for instance to represent the trace of a transition system.
Such an object is a potentially infinite sequence of \textit{internal events}, \Tau,
and \textit{external} (or \textit{visible}) events $\Vis{n}$, terminated (if
finite) by the \Ret marker. Here, for simplicity, we assume that visible events
carry a natural number. We will sometimes omit the $\beta$ constructor and just
write $n$ (especially in examples) to save space.

Here are some example streams:
\[
  \begin{array}{lll}
    s_0 = & \Viss{0} \app \Viss{1} \app \Ret & \mbox{finite stream}
    \\
    s_1 = & \Tau \app \Viss{0} \app \Tau \app \Tau \Viss{1} \app \Ret  &  \mbox{finite stream}
    \\
    s_2 = & \Viss{0} \app \Viss{1} \app \Viss{2} \app \ldots \app \Viss{n} \app \Viss{(n+1)} \app \ldots & \mbox{infinite increasing stream}
    \\
    s_3 = & \Viss{0} \app \Tau \app \Viss{1} \app \Tau \app \Viss{2} \app \ldots \app \Viss{n} \app \Tau \app \Viss{(n+1)} \app \ldots & \mbox{infinite increasing stream}
    \\
    s_4 = & \Viss{0} \app \Viss{1} \app \Viss{0} \app \Viss{1} \app \Viss{0} \app \Viss{1} \app \Viss{0} \app \Viss{1} \app \ldots & \mbox{infinite alternating stream}
    \\
    s_5 = & \Tau \app     \Tau \app     \Tau \app     \Tau \app     \Tau \app     \Tau \app     \Tau \app     \ldots & \mbox{silent divergence}

  \end{array}
\]

It is well-known that strong bisimulation is often too tight a relation to be
relevant when studying such systems. One should instead work ``\utt,''
which means that, when considering whether two streams are ``the same,'' we can
disregard any finite number of \Tau steps on either side. This \textit{weak
bisimulation} matches terminal constructors and identical external events
one-to-one, but also allows for a finite number of \Tau steps to be stripped
away from either stream at any given point.  We write $s \eutt t$ to mean that $s$ is equivalent to $t$ \utt (which we often abbreviate to $\mathtt{eutt}$).
For the examples shown above, we have $s_0 \eutt s_1$ and $s_2 \eutt s_3$, but no other distinct pairs of streams are weakly bisimilar.

We delay the full exposition of a formal definition of this relation to
Section~\ref{sec:bisim}. Here, we simply observe that we can define $\eutt$ as
the greatest fixed point of a functor, \euttF:
\[
  \begin{array}{c}
    \euttF:~\mathcal P (\mathtt{stream} \times \mathtt{stream}) \rightarrow \mathcal P (\mathtt{stream} \times \mathtt{stream})
    \\
    \eutt \ \equiv \gfp{\euttF}
  \end{array}
\]

\noindent We can think of \euttF{} as acting on a set of pairs of streams $Y$,
which behaves as the ``coinductive hypothesis'' in this definition. \euttF{} is
defined so that it satisfies several properties that characterize weak
bisimulation. Among them, we have:

\begin{lemma}[\euttF{} Tau Left] \label{lemma:euttf-tau-left}
  \[ X \subseteq \euttF(Y) \implies \{(\Tau \app s, t) \ | \ (s,t) \in X\} \subseteq \euttF(Y) \]
\end{lemma}

\begin{lemma}[\euttF{} Vis]\label{lemma:euttf-vis}
  \[ X \subseteq Y \implies \{(\Viss{n} \app s,\Viss{n} t) \ | \ (s,t) \in X\} \subseteq \euttF(Y) \]
\end{lemma}

\noindent The first lemma states that, when reasoning backwards using
goal-directed proof search, if we want to show that $\Tau{}\cdot s$ is related to $t$
by $\euttF(Y)$, it suffices to show that $s$ is related to $t$ by
$\euttF(Y)$---we can drop a \Tau from the left stream. The second lemma states
that if two streams begin with the same visible event $n,$ we can directly
appeal to the coinductive hypothesis $Y$ to establish the relation.

With this setup, we can give an example proof using \pacon-style reasoning and
see where it can be improved upon.

Consider the two transition systems $s$ and $t$ depicted in
Figure~\ref{fig:ex-ts}.
They each visually encode the different states two streams can be in.
A stream can change state through either an internal step or by emitting an event.
We also consider additional equations we
know over the states of the streams: the edge labeled by an equality sign
represents definitional equality -- we assume we have such an equation in our
context.
The bottom half of Figure~\ref{fig:ex-ts} characterizes the same two streams, but
as a system of equations.

\begin{figure}[t]
  \begin{minipage}{0.23\textwidth}
    \centering
    \begin{tikzpicture}[->, auto, scale=0.8, node distance=1.5cm]
      \tikzset{every state/.style={minimum size=0.5em}}

      \node[state] (t0)                {$s_0$};
      \node[state] (t'0) [below of=t0] {$s'_0$};

      \node[state] (t1) [right of=t0] {$s_1$};
      \node[state] (t'1) [below of=t1] {$s'_1$};

      \path
      (t0) edge [left] node {$\Viss{0}$} (t'0)
      (t'0) edge [sloped] node {$\Tau$} (t1)

      (t1) edge node {$\Viss{1}$} (t'1)
      (t'1) edge node {$\Viss{2}$} (t'0);
    \end{tikzpicture}
  \end{minipage}
  \begin{minipage}{0.23\textwidth}
    \centering
    \begin{tikzpicture}[->, auto, scale=0.8, node distance=1.5cm]
      \tikzset{every state/.style={minimum size=0.5em}}

      \node[state] (t0)                {$t_0$};
      \node[state] (t'0) [below of=t0] {$t'_0$};

      \node[state] (t1) [right of=t0] {$t_1$};
      \node[state] (t'1) [below of=t1] {$t'_1$};

      \path
      (t0) edge [left] node {$\Viss{0}$} (t'0)
      (t'0) edge [-, sloped] node {$=$} (t1)

      (t1) edge node {$\Viss{1}$} (t'1)
      (t'1) edge node {$\Viss{2}$} (t'0);
    \end{tikzpicture}
  \end{minipage}
  \[
  \begin{array}{@{}r@{\;}l@{\qquad}r@{\;}l@{\qquad}r@{\;}l@{\qquad}r@{\;}l@{}}
    s_0 &= \Viss{0}\, s'_0 &
    s'_0 &= \Tau\, s_1 &
    s_1 &= \Viss{1}\, s'_1 &
    s'_1 &= \Viss{2}\, s'_0
    \\
    t_0 &= \Viss{0}\, t'_0 &
    t'_0 &= t_1 &
    t_1 &= \Viss{1}\, t'_1 &
    t'_1 &= \Viss{2}\, t'_0
  \end{array}
  \]
  \caption{Two weakly bisimilar transition systems: illustrating the shortcoming of \pacon's reasoning principles}
  \label{fig:ex-ts}
\end{figure}
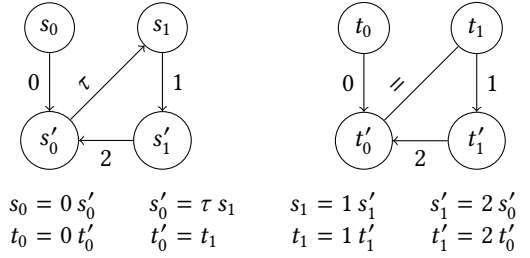

\begin{figure}[t]
Let $X_0 = \{(s_0,t_0), (s_1,t_1)\}$ and
$X_1 = \{(s_0',t_0'), (s_1',t_1')\}$
\begin{align*}
  & X_0 \subseteq \gfp{\euttF}
  \iffv{\text{\textsc{Init}}} X_0 \subseteq \paco{\euttF}{\emptyset}
  \\[-4pt]
  \iffv{\text{\textsc{Acc} \textbf{(a)}}}\quad &X_0 \subseteq \paco{\euttF}{X_0}
  \\[-4pt]
  \iffv{\text{\textsc{Unfold}}}\quad& X_0 \subseteq \euttF(X_0 \cup \paco{\euttF}{X_0})
  \\[-4pt]
  \impliedbyv{\text{lem.~\ref{lemma:euttf-vis} \textbf{(b)}}}\quad &X_1 \subseteq X_0\cup\paco{\euttF}{X_0}
  \\[-2pt]
  \impliedby\quad & X_1 \subseteq \paco{\euttF}{X_0}
  \\[-4pt]
  \iffv{\text{\textsc{Acc} \textbf{(c)}}}\quad & X_1 \subseteq \paco{\euttF}{(X_0\cup X_1)}
  \\[-3pt]
  &\mysmall{\text{We now handle both cases in $X_1$ separately:}}
  \\[-3pt]
  \mysmall{\textbf{rhs:}}~\quad& (s_1',t_1')\in \paco{\euttF}{(X_0\cup X_1)}
  \\[-4pt]
  \iffv{\text{\textsc{Unfold}}}\quad& (s_1',t_1')\in  \euttF(X_0 \cup X_1 \cup \paco{\euttF}{(X_0 \cup X_1)})
  \\[-4pt]
  \impliedbyv{\text{lem.~\ref{lemma:euttf-vis}}}\quad &\specialcell{(s_0',t_0')\in (X_0 \cup X_1 \cup \paco{\euttF}{(X_0 \cup X_1)})\hfill\qed}
  \\[-1pt]
  \mysmall{\textbf{lhs:}}~\quad& (s_0',t_0')\in  \paco{\euttF}{(X_0\cup X_1)}
  \\[-3pt]
  &\mysmall{\underline{Solution with redundancy \textbf{(d)}:}}
  \\[-5pt]
  \iffv{\text{\textsc{Unfold}}}\quad& (s_0',t_0')\in  \euttF(X_0 \cup X_1 \cup \paco{\euttF}{(X_0 \cup X_1)})
  \\[-4pt]
  \impliedbyv{\text{lem.~\ref{lemma:euttf-tau-left};~\ref{lemma:euttf-vis}}}\quad &\specialcell{(s_1',t_1')\in (X_0 \cup X_1 \cup \paco{\euttF}{X_0 \cup X_1})\hfill\qed}
  \\[-3pt]
  &\mysmall{\underline{Failed attempt without redundancy \textbf{(e)}:}}
  \\[-5pt]
  \impliedbyv{\text{lem.~\ref{lemma:paco-clo-tau-left}}}\quad &(s_1,t_1)\in \paco{\euttF}{(X_0\cup X_1)}\mysmall{ : we cannot conclude.}
\end{align*}
\caption{Shortcoming of \pacon: an illustrating proof}
\label{fig:paco-short}
\end{figure}

Their behaviors can therefore be described as follows. Both streams
consist of an infinite cycle alternating between the visible events
$1$ and $2$. In the left stream, each iteration of these two events
is separated by a silent step, while the right stream starts the new cycle
immediately---embodied by the definitional equality between $t_0'$ and $t_1$.
Finally, both streams have an initial state stepping into the cycle by emitting
$0$.

We wish to build a weak bisimulation between both corresponding upper states of
$s$ and $t$, that is to prove that $s_0 \eutt t_0$ and $s_1 \eutt t_1$. The \pacon library
is the perfect tool for such a task: we would like to build our proof incrementally as
we explore the underlying transition systems. Let us venture step by step into this task,
depicted in Figure~\ref{fig:paco-short}.

This minimal example highlights a deep problem in the existing reasoning principles:
unused accumulated knowledge is always guarded again, i.e. sent back behind the guard.
We see this in the proof 
at the point where we use \textsc{Acc} for the second time (marked \textbf{(c)}).
We had already used \textsc{Acc} once, at point \textbf{(a)}, putting $X_0$ into the accumulated knowledge.
Intuitively, this means that after we step under a guard we should be able to use $X_0$, which is what happens at point \textbf{(b)}, where we have $X_0$ directly available on the right hand side.
The problem is that even though the knowledge $X_0$ is available at point \textbf{(b)}, we have to discard it to use \textsc{Acc} at point \textbf{(c)}, which forgets the fact that $X_0$ was available.

The impact of this loss of information shows up later, when trying to conclude
for the pair of states $(s_0',t_0')$. A natural solution, depicted at point
\textbf{(d)}, is to simply blindly go through a new round of unfolding and
stepping under the functor, using Lemmas~\ref{lemma:euttf-tau-left}~and~\ref{lemma:euttf-vis} successively.
Note that Lemma~\ref{lemma:euttf-tau-left} alone is not enough to go under the functor,
it does not act as a guard.
\ph{Maybe the terminology needs to be changed here}
However, by taking this step, we are repeating a part of the proof
we already did: taking the transition that emits a $1$ for both streams. This
may seem innocuous on such a toy example, but may in general require reiterating
an arbitrarily complex proof.

Performing the case analysis earlier (or proving the equivalence of different states)
would have avoided the issue with repeated reasoning in this case.
However, this solution is both cumbersome and not always possible.
For example, the more complex data type described in Section~\ref{sec:itrees} has
a branching structure that renders such solutions ineffective.

Intuitively however, we would like to simply ignore this $\tau$ on $s$ and conclude by using $X_0$, knowledge that we made available earlier in the proof.
The first part of this intuition, \ph{mention the second part later?} the innocuousness of the $\tau$ guard,
is a particular case of a more general reasoning principle:
reasoning \emph{up-to} silent steps. We can indeed formalize this idea using $\pacon$,
by proving the following lemma:

\begin{lemma}[$\paco{\euttF}{}$ Tau Left] \label{lemma:paco-clo-tau-left}
  \[ X \subseteq \paco{\euttF}{(Y)} \implies \{(\Tau \app s, t) \ | \ (s,t) \in X\} \subseteq \paco{\euttF}{(Y)} \]
\end{lemma}

It precisely states that one can strip a $\tau$ from the left hand side under a
call to $\paco{\euttF}{}.$ Using this lemma at point \textbf{(e)} in
Figure~\ref{fig:paco-short}, we can therefore reduce our goal to relating
the desired pair, $(s_1,t_1).$ However this is useless in this case due to
\pacon's inability to remember previously available knowledge in the presence of
nested accumulation lemmas: we
know that the pair of states are in $X_0$, knowledge that was made available before,
and yet we cannot access it to conclude.

To alleviate these difficulties, we introduce a new construction that still supports
up-to reasoning, but crucially offers a finer grained management of available knowledge.

\section{Generalized Parameterized Coinduction}
\label{sec:gpaco}
In this paper, we introduce a new construct, dubbed the \emph{generalized
  parameterized greatest fixed point} (and succinctly referred to as \gpacon), that we show satisfies new
principles that greatly ease reasoning in cases such as the one depicted in
Figure~\ref{fig:ex-ts}. Our new construct builds on the so-called parameterized
greatest fixed point introduced by \citeN{paco}, and implemented in Coq
through the \pacon library.

We extend the parameterized greatest fixed point in two ways.  First, we refine its treatment of available knowledge by making a distinction between knowledge that is available, or ``already unlocked,'' and knowledge that is guarded, or ``must be unlocked.'' Maintaining this distinction dramatically simplifies incremental coinductive proofs.  Second, we build in support for ``up-to'' reasoning, another powerful technique that lets us construct coinductive relations using closure operators.

\subsection{Generalized Incremental Reasoning}
Recall our unsatisfactory proof in Figure~\ref{fig:paco-short}.
One core issue comes from the fact that while the accumulated knowledge is
safely released after a guard, it does not internalize the fact that this
knowledge became available. The first extension we introduce is to precisely
take this observation into account: the parameterized greatest fixed point is now
parameterized by \emph{two} elements representing accumulated knowledge.

The generalized parameterized greatest fixed point $\gpaconoclo {f} {r} {g}$,
also shortened to \gpacon, therefore
intuitively represents the greatest fixed point of the functor $f$ with \emph{available}
accumulated knowledge $r$ and \emph{guarded} accumulated knowledge $g$, which becomes available only after making progress by applying $f$.
We express this distinction in the following definition, which uses $\paco{f}{-}$.

\begin{definition}[Generalized parameterized greatest fixed point (first definition)]
  ~\\Define $\hat G \in (C \mon C) \mon (C \mon C \mon C)$ to be:
  $$\gpaconoclo {f} {r} {g} \defeq r \sqcup \paco {f} {(r \sqcup g)}$$
\end{definition}

Note that if we pick $r = \bot$, this definition degenerates to $\paco {f} {g}$, which gives us the following soundness property.
As before, we call it \textsc{Init} because it lets us begin a coinductive proof by moving into the \gpacon realm.
\footnote{We overload the lemma names like \textsc{Init} and \textsc{Acc} which are defined both for $\paco{f}{-}$ and $\gpaconoclo{f}{-}{-}$.  Which one is meant can easily be distinguished from the context.}

\begin{lemma}[\textsc{Init}]
  $$\gpaconoclo {f} {\bot} {\bot} \equiv \paco {f} {\bot} \equiv \gfp{f}$$
\end{lemma}

We can also return to vanilla parameterized coinduction from the generalized version:
\begin{lemma}[\textsc{Final}]
  $$ r \sqcup \paco {f} {g} \sqsubseteq \gpaconoclo {f} {r} {g} $$
\end{lemma}
These two lemmas mean in particular that \gpacon is fully backwards compatible
with \pacon: no changes in previous definitions or statements written with \pacon are required,
and the new reasoning principles are available for properties defined in terms of $\paco{}$.

The \textsc{Base} equation below embodies the fact that available knowledge is stored
in \gpacon. By definition, it is indeed trivial to see that $r$ is immediately available for use:
\begin{lemma}[\textsc{Base}]
  $$r \sqsubseteq \gpaconoclo {f} {r} {g}$$
\end{lemma}

Naturally, in order for \textsc{Base} to be sound, the incremental principle extends only the guarded knowledge:
\begin{lemma}[\textsc{Acc}]
  $$x \sqsubseteq \gpaconoclo {f} {r} {(g \sqcup x)} \iff x \sqsubseteq \gpaconoclo {f} {r} {g} $$
\end{lemma}

Finally, stepping under the guard makes the guarded knowledge available. Note
that the pattern of accumulation ensures that we always have the invariant that
$r \sqsubseteq g$, which is why erasing $r$ here does not lose information.

\begin{lemma}[\textsc{Step}]
  $$f(\gpaconoclo {f} {g} {g}) \sqsubseteq \gpaconoclo {f} {r} {g} $$
\end{lemma}

With the addition of the available knowledge parameter to \gpacon and its new reasoning principles, we are closer to a more succinct proof for Figure~\ref{fig:ex-ts} without the extraneous steps required in the previous proof.
However, we still need a statement analogous to Lemma~\ref{lemma:paco-clo-tau-left}, in order to strip off a \Tau without having to continue to go under guards.

\begin{lemma}[$\gpaconoclo{\euttF}{}{}$ Tau Left, idealized] \label{lemma:gpaco-clo-tau-left}
  \[ X \subseteq \gpaconoclo{\euttF}{r}{g} \implies \{(\Tau \app s, t) \ | \ (s,t) \in X\} \subseteq \gpaconoclo{\euttF}{r}{g} \]
\end{lemma}

Note that this lemma \emph{does not} hold with the definition of \gpacon introduced in this subsection.
We will get back to its proper statement, as well as its soundness, in
Section~\ref{sec:up-to}, once we have extended \gpacon with intrinsic support
for up-to reasoning. Accepting temporarily this slight idealization,
we showcase in Figure~\ref{fig:paco-ideal} a proof of the example from
Section~\ref{sec:background} which eliminates the undesired repetition.

\begin{figure}
\begin{align*}
  & X_0 \subseteq \gfp{\euttF}
  \iffv{\text{\textsc{Init}}} X_0 \subseteq \gpaconoclo{\euttF}{\emptyset}{\emptyset}
  \\[-4pt]
  \iffv{\text{\textsc{Acc} \textbf{(a)}}}\quad &X_0 \subseteq \gpaconoclo{\euttF}{\emptyset}{X_0}
  \\[-4pt]
  \impliedbyv{\text{\textsc{Step} \textbf{(b)}}}\quad& X_0 \subseteq \euttF(\gpaconoclo{\euttF}{X_0}{X_0})
  \\[-4pt]
  \impliedbyv{\text{lem.~\ref{lemma:euttf-vis}}}\quad &X_1 \subseteq \gpaconoclo{\euttF}{X_0}{X_0}
  \\[-4pt]
  \iffv{\text{\textsc{Acc} \textbf{(c)}}}\quad & X_1 \subseteq \gpaconoclo{\euttF}{X_0}{(X_0\cup X_1)}
  \\[-4pt]
  \mysmall{\textbf{rhs:}}\quad & (s_1', t_1') \in \gpaconoclo{\euttF}{X_0}{(X_0\cup X_1)}
  \\[-4pt]
  \impliedbyv{\text{\textsc{Step}}}\quad & (s_1', t_1') \in \euttF (\gpaconoclo{\euttF}{(X_0\cup X_1)}{(X_0\cup X_1)})
  \\[-4pt]
  \impliedbyv{\text{lem.~\ref{lemma:euttf-vis}}}\quad & (s_0', t_0') \in \gpaconoclo{\euttF}{(X_0\cup X_1)}{(X_0\cup X_1)}
  \\[-4pt]
  \impliedbyv{\text{\textsc{Base}}}\quad & \specialcell{(s_0', t_0') \in X_0\cup X_1 \hfill \qed}
  \\[-4pt]
  \mysmall{\textbf{lhs:}}\quad & (s_0', t_0') \in \gpaconoclo{\euttF}{X_0}{(X_0\cup X_1)}
  \\[-4pt]
  \impliedbyv{\text{lem.~\ref{lemma:gpaco-clo-tau-left}}}\quad & (s_1, t_1) \in \gpaconoclo{\euttF}{X_0}{(X_0\cup X_1)}
  \\[-4pt]
  \impliedbyv{\text{\textsc{Base} \textbf{(d)}}}\quad & \specialcell{ (s_1, t_1) \in X_0 \hfill \qed}
\end{align*}
\caption{Improved proof for Figure~\ref{fig:ex-ts}}
\label{fig:paco-ideal}
\end{figure}

This proof illustrates how the extra parameter provides just the right degree of freedom to remember knowledge collected across nested calls to \textsc{Acc}.
Here, the first use of \textsc{Acc} at point \textbf{(a)} doesn't yet provide any more flexibility compared to the old proof.
At point \textbf{(b)}, however, the \textsc{Step} operation copies $X_0$ from the ``guarded knowledge'' parameter to the ``available immediately'' parameter.
Later, at the second use of \textsc{Acc} at point \textbf{(c)}, $X_0$ remains available, even as $X_1$ is placed under the guard.
The payoff comes at point \textbf{(d)}, where we can immediately use $X_0$.

This example shows how the additional parameter allows for smoother reasoning and less redundancy in the proofs.
One might wonder: are two parameters enough?
Might we need an even more general version with three or four parameters to use in some other proof?
The answer is that no, two are sufficient.
Intuitively, any particular fact is either available or still guarded.
The two parameters partition the knowledge into those categories, and the lemmas manipulate the knowledge precisely.

\subsection{Up-to Reasoning: Generalized \pacon{} with Closure}
\label{sec:up-to}

The ability to construct coinductive proofs incrementally, as considered above,
is one technique that is invaluable for working with coinduction in an automated
theorem prover. Another crucial technique is the use of ``up-to'' reasoning
principles, which enable more scalable and modular proofs.

The basic idea is to define a closure operator $\mathit{clo} \in C \to C$ that,
given a relation $X$, extends it to a larger relation $\mathit{clo}(X)$.
Then such an up-to technique $\mathit{clo}$ allows us to work with smaller
relations when proving, for example, bisimilarity, reducing the effort required in the proof.
The power of an up-to technique lies in the fact that the smaller relation $X$ may not
be a bisimulation at all.
However, for reasoning up-to $\mathit{clo}$ to be sound, $X$ must be contained in a bisimulation.
For a more in-depth description of up-to techniques, see \cite{pous_sangiorgi_2011}.

For example, the closure operator used for Lemma~\ref{lemma:gpaco-clo-tau-left} is:
\[ \tau_L(R) = \{ (\Tau^* s, t) \ | \ (s,t) \in R \} \]
where $\Tau^*$ means any finite number of $\Tau$s.
Using this up-to technique frees the user from having to manually step through the functor
and build the bisimulation relation by manipulating $\Tau$s one by one on the left side.
In this section, we develop the enhancements to \gpacon necessary
to reason using these closure operators.

Before we proceed, we briefly review the state-of-the-art up-to techniques.
\citeN{Pous16} characterizes valid closures as any
function bounded by the greatest \emph{compatible} closure, called
the \emph{companion}. Specifically, an up-to function $\mathit{clo} \in C \mon C$
is \emph{compatible} with $f$ if $\mathit{clo} \circ f \sqsubseteq f \circ \mathit{clo}$.
Compatible functions are a class of up-to techniques that are nice to work with because they are compositional, so different compatible up-to techniques can be used in a single proof.
The companion $\cpn_f \in C\mon C$ is the join of all
such compatible functions, which is again compatible with $f$.
Then, $\cpn_f$ admits nice incremental and up-to principles for coinduction:
in particular, $\mathit{clo} (\cpn_f (r)) \sqsubseteq \cpn_f(r)$ for any
(not necessarily compatible) function $\mathit{clo} \sqsubseteq \cpn_f$.
In practice, most useful up-to functions are bounded by the companion.

In our approach, instead of using the companion, we parameterize our construct with the
upper bound of valid closures, which we call a \emph{base closure}, in
order to allow a more explicit construction of the fixed point. This
generalization is essential in the development of our
equational theory for weak bisimulation in Section~\ref{sec:euttG}.
\begin{definition}[Generalized parameterized greatest fixed point]
  We redefine the previous $\hat G$,
  adding the base closure $\bclo \in C\mon C$ as the second argument:
  $$\gpaco {f} {\bclo} {r} {g} \defeq \bclo^* (r \sqcup \paco{f \circ \bclo^*} {(r \sqcup g)} )$$
  where $\bclo^*$ is the transitive closure of $\bclo$.
\end{definition}
Note that by choosing the companion as a base closure,
we get the equality $\gpaco{f}{\cpn_f}{r}{g} = \cpn_f (r \sqcup f(\cpn_f (r \sqcup g)))$.

\begin{definition}
  We introduce the following useful notation:
$$\gupaco {f} {\bclo} {g} \defeq \gpaco {f} {\bclo} {g} {g}$$
\end{definition}

Then we can use any up-to function $\mathit{clo}$ bounded by $\bclo$, and in fact even larger ones bounded by $\gupaco {f} {\bclo} {}{}$.
\begin{lemma}[\textsc{Closure}]
  \label{lem:clo}
  If $\mathit{clo} \sqsubseteq \gupaco {f} {\bclo} {}{}$, then
  $$ \mathit{clo} (\gpaco {f} {\bclo} {r} {g}) \sqsubseteq \gpaco {f} {\bclo} {r} {g}$$
\end{lemma}
Since $\bclo \sqsubseteq \gupaco {f} {\bclo} {}{}$, in the case $\mathit{clo} = \bclo$, it is always valid to use \textsc{Closure}, which will be marked as \textsc{Closure*}.

Using this rule, we can now amend Lemma~\ref{lemma:gpaco-clo-tau-left}: it holds, provided we instantiate $\bclo$ with $\tau_L$ or another closure that contains it (in the sense of Lemma~\ref{lem:clo}).
For the overall approach to be sound, the usual criterion required of such a base closure is a notion of compatibility.
We work with a relaxed condition, weak compatibility, that can be seen as an instance of a compatible up-to-function function~\cite{Pous16}:
\begin{definition}[Weakly compatible closure]
  $\bclo \in C \mon C$ is weakly compatible for $f$ if
  $$\bclo \circ f \sqsubseteq f \circ \gupaco {f} {\bclo} {}$$
\end{definition}

We can begin using generalized parameterized coinduction from usual parameterized coinduction:
\begin{lemma}[\textsc{Init}]
  \label{lem:init}
  If $\bclo$ is weakly compatible for $f$, then
  $$\gpaco {f} {\bclo} {\bot} {\bot} \sqsubseteq \paco {f} {\bot}$$
\end{lemma}

\begin{figure}
  \[
  \begin{array}{@{}r@{\;}l@{\qquad}r@{\;}l@{\qquad}r@{\;}l@{\qquad}r@{\;}l@{}}
    s_0 &\bisim \Viss{0}\, s'_0 &
    s'_0 &\bisim r \cat s_1 &
    s_1 &\bisim \Viss{1}\, s'_1 &
    s'_1 &\bisim \Viss{2}\, s'_0
    \\
    t_0 &\bisim \Viss{0}\, t'_0 &
    t'_0 &\bisim r' \cat t_1 &
    t_1 &\bisim \Viss{1}\, t'_1 &
    t'_1 &\bisim \Viss{2}\, t'_0
  \end{array}
  \]
  \caption{Two weakly bisimilar streams when $r\eutt r'$}
  \label{fig:ex-upto}
\end{figure}

\begin{figure*}
  \begin{centermath}
    \infer[\small{\textsc{Init}}]
          {\gpaco {f} {\bclo} {\bot} {\bot} \sqsubseteq \paco {f} {\bot}}
          {\small{\bclo \text{ weakly compatible for } f}}
    \hfill
    \infer[\small{\textsc{Base}}]
          {r \sqsubseteq \gpaco {f} {\bclo} {r} {g}}{}
    \hfill
    \infer[\small{\textsc{Final}}]
          {r \sqcup \paco {f} {g} \sqsubseteq \gpaco {f} {\bclo} {r} {g}}{}
    \hfill
    \infer[\small{\textsc{Step}}]
          {f(\gpaco {f} {\bclo} {g} {g}) \sqsubseteq \gpaco {f} {\bclo} {r} {g}}{}
  \end{centermath}
  \begin{centermath}
    \hfill
    \infer[\small{\textsc{Acc}}]
          {x \sqsubseteq \gpaco {f} {\bclo} {r} {g}}
          {x \sqsubseteq \gpaco {f} {\bclo} {r} {(g \sqcup x)}}
    \hfill
    \infer[\small{\textsc{Closure}}]
          {clo (\gpaco {f} {\bclo} {r} {g}) \sqsubseteq \gpaco {f} {\bclo} {r} {g}}
          {clo \sqsubseteq \gupaco {f} {\bclo} {}}
    \hfill
    \infer[\small{\textsc{Closure*}}]
          {\bclo (\gpaco {f} {\bclo} {r} {g}) \sqsubseteq \gpaco {f} {\bclo} {r} {g}}
          {}
    \hfill
  \end{centermath}

  \caption{Proof rules for generalized parameterized coinduction}
  \label{fig:gpaco-axiom}
\end{figure*}

For a more involved example showing how reasoning up-to closures can help, consider the streams in
Figure~\ref{fig:ex-upto}, which are a modified version of the example we saw earlier in Figure~\ref{fig:ex-ts}.
Here, rather than $s$ taking an extra $\Tau$ step, both streams go through intermediate transitions $r$ and $r'$ respectively.
Moreover, rather than defining the streams using definitional equality ``$=$'', we instead specify them via strong bisimilarity ``$\bisim$''.
In the case that $r$ and $r'$ are known to be weakly bisimilar to each other, the resulting streams remain weakly bisimilar.
However, in order to prove that this is the case, the weak bisimulation relation would have to
contain all of the internal bisimilar states of $r$ and $r'$, and moreover, it
would have to somehow incorporate the states related by the underlying strong
bisimilarity relation too.

\ph{Similarly to what? feels like we're missing a paragraph here}
Similarly, when proving equivalence \utt, it is intuitively
the case that if $r \eutt r'$ and we want to coinductively relate the concatenated streams $r \cat s \eutt r' \cat t$, it suffices to relate $s$ and $t$---we can ignore the weakly
bisimilar prefixes and focus on proving the tails of the streams equivalent.

Up-to reasoning formalizes these intuitions.
First, we define two closure operators, up-to prefix and up-to (strong) bisimilarity:
\[ \mathit{prefix}(R) = \{ (h_1 \cat t_1, h_2 \cat t_2) \mid h_1 \eutt h_2 \land (t_1, t_2) \in R \}\]
\[ \mathit{bisim}(R) = \{ (a, b) \mid \exists a', b', a \bisim a' \land b \bisim b' \land (a', b') \in R \} \]

\noindent Being able to prove $s_0 \eutt t_0$ and $s_1 \eutt t_1$ up-to $\mathit{bisim}$ and
$\mathit{prefix}$ allows for a proof conducted parametrically in the assumption
$r \eutt r'$, leading to a proof with complexity similar to the one for
Figure~\ref{fig:ex-ts}.
Note that up-to $\mathit{prefix}$ is an instance of the standard up-to context technique~\cite{pous_sangiorgi_2011}.

Using the resulting set of reasoning principles provided by \gpacon, summarized
in Figure~\ref{fig:gpaco-axiom}, we can proceed with the proof of weak
bisimilarity for Figure~\ref{fig:ex-upto}, that is $s_0 \eutt t_0$ and $s_1 \eutt t_1$.
We use $\mathit{bisim}$ as our base closure, a choice that will be grounded \ph{wording?} in Section~\ref{sec:bisim}.

By leveraging the reasoning principles of up-to $\mathit{bisim}$ and
$\mathit{prefix}$, we can derive a proof extremely similar to the previous examples.
The difference lies in the application of the \textsc{Closure} rules at five points in the proof.
We first apply \textsc{Closure*} twice with $\mathit{bisim}$ to rewrite $s_0$, $t_0$, $s_1$, and $t_1$.
Next we apply \textsc{Closure*} again to replace $s_0'$ and $t_0'$ with $r \cat s_1$ and $r'_n \cat t_1$ respectively.
We then apply \textsc{Closure} with $\mathit{prefix}$ to remove the weakly bisimilar prefixes $r$ and $r'$.
Finally we apply \textsc{Closure*} with $\mathit{bisim}$ again to rewrite $s_1'$ and $t_1'$.
The remainder of the proof follows as before.


\section{Up-to-tau Bisimulation of Streams}
\label{sec:bisim}

In the previous section we introduced \gpacon, a greatest fixed point
predicate recording both the accumulated knowledge guarded by a constructor
and its already accessible counterpart. We additionally extended the
construction to internalize the support for up-to closure.

We have described the novel, richer reasoning principles derived from \gpacon.
We now illustrate its practical use concretely by establishing
a rich equational theory to reason about weak bisimilarity of interactive systems.
We develop this case study using the data type of potentially infinite streams of
internal and external events, and study their equivalence up to internal steps.

The approach and the results being general, we present them in lattice theoretic
notations, but all results are formalized in Coq.

\subsection{Streams}

The data type considered is the same type of potentially finite streams of internal
and external events introduced earlier in the paper. Formally, we define
$\mathtt{stream} \defeq \gfp{\mathtt{streamF}}$ where:
\begin{align*}
  \mathtt{streamF}~X~ \defeq
  \{\Ret\} &\cup
  \{\Tau\cdot s~\mid~s\in X \} \\
  &\cup \{\Vis{n}\cdot s~\mid~s\in X,~n\in\mathbb N\}
\end{align*}

An element of the resulting type \texttt{stream}
is hence a potentially infinite trace consisting of internal steps, represented as $\tau$
constructors, and visible events, emitting natural numbers, represented as
$\beta$ constructors. Such a data type can for instance be thought of as the
observable trace of an interactive program's execution.

We fix the lattice of interest to $\mathcal P(\mathtt{stream} \times \mathtt{stream})$ in the rest of the paper.

Defining a concatenation operation over streams, \texttt{concat}, is straightforward: let
$\mathtt{concat} \defeq \gfp{\mathtt{concatF}}$ where
\[\arraycolsep=1.4pt\begin{array}{ll}
  \mathtt{concatF}~concat\_~ \defeq \lambda s~k.&\mathtt{case}~s~\mathtt{of}\\
  & \mid~\Ret~\Rightarrow~k\\
  & \mid~\Tau\cdot s~\Rightarrow~\tau\cdot (concat\_~s~k)\\
  & \mid~\Vis{n}\cdot s~\Rightarrow~\Vis{n}\cdot (concat\_~s~k)
\end{array}
\]
We write $s\cat t$ for $\mathtt{concat}~s~t$.

Reasoning about these streams naturally requires to prove that \texttt{concat}
respects an equivalence relation over streams, which justifies reasoning
principles such as: $s \eutt t \implies s \cat k \eutt t \cat k$.
The usual notion of Leibniz equality is inadequate when manipulating coinductive types.
Instead, the standard equivalences used to reason about such streams are the
notions of strong and weak bisimulations.

\subsection{Bisimulation, Equivalence Up-to-tau}

A natural equivalence relation over stream is to require the shape of both
streams to match exactly, systematically pairing the head constructors.
This coinductive relation, known as \emph{strong bisimulation},
is convenient to work with, but too restrictive in practice.
Indeed, it not only observes the visible events two systems emit when
comparing them, but also ensures that their internal steps match as well:
in a sense, it is a timing-sensitive equivalence of processes.

\emph{Equivalence up-to-tau} is a form of weak bisimulation, a coarser
relation than strong bisimulation. It ignores any finite amount of internal
steps a process may take before reaching its next external event. This relation
is much more useful in practice, and is notably the de facto standard used in
verified compilation to express the semantic preservation
criterion~\cite{compcert,cakeml}.

Equivalence up-to-tau has to be careful not to relate the infinite
sequence of $\tau$ with all streams. This is achieved by an inductive-coinductive
definition: the functor \texttt{bisimF} whose greatest fixed point we take is
itself defined recursively, but as a smallest fixed point. This nested structure
makes it particularly delicate to work with without a carefully crafted
metatheory.  Moreover, because strong and weak bisimilarity have some
common structure, it is beneficial for proof engineering purposes to share
as much of their common metatheory as possible.

We demonstrate in this section how introducing a parameterized
version of the weak bisimulation relation allows us to derive a rich equational
theory that alleviates the pain of working with nested inductive-coinductive
definitions. Our new construction, \gpacon, is instrumental to the proofs in
this theory.

\subsection{A Family of Bisimulations}

\begin{figure}
  \begin{flalign*}
  \mathtt{fix}~&\mathtt{bisimF}~(b_L~b_R: \mathtt{bool})~\betaclo~X \defeq \\
  &\{(\Ret,\Ret)\} ~\cup\\
  &\{(\tau\cdot s,\tau\cdot t)~\mid~(s,t)\in X \} ~\cup\\
  &\{(\Vis{n}\cdot s,\Vis{n}\cdot t)~\mid~(s,t)\in \betaclo(X),~n\in\mathbb N\}~\cup\\
  &\{(\tau\cdot s, t)~\mid~ b_L=\btrue \land (s,t)\in \mathtt{bisimF}~b_L~b_R~\betaclo ~X \}~ \cup\\
  &\{(s, \tau\cdot t)~\mid~ b_R=\btrue \land (s,t)\in \mathtt{bisimF}~b_L~b_R~\betaclo ~X \}\\
  \multispan2{$\mathtt{bisim}~b_L~b_R \defeq~ \paco{\mathtt{bisimF}~b_L~b_R~\mathtt{id}}{\bot}$\hfil}
  \end{flalign*}
  \caption{Definition of a family of bisimulations over streams
  }
  \label{fig:bisimgen}
\end{figure}

While weak bisimulation is the core relation we care about, several related
relations are relevant to prove our equational theory. As a way to factor work,
we start by defining in Figure~\ref{fig:bisimgen} \texttt{bisim}, a family of relations
over streams. Let us for now ignore its three parameters and focus at a high
level on the functor \texttt{bisimF \_ \_ \_ X}. We use the \texttt{fix} keyword
as a notation to express \texttt{bisimF} itself is defined as a smallest fixed point.

There are five ways we may relate two streams:
\begin{enumerate*}
\item by matching \Ret constructs,
\item by matching $\tau$ and co-recursing,
\item by matching identical $\beta$ and co-recursing,
\item by stripping a $\tau$ from the left and recursing or
\item by stripping a $\tau$ from the right and recursing.
\end{enumerate*}
Note the use of a \textit{recursive} call when stripping $\tau$ in the
asymmetric cases (4) and (5):
if we were to iterate co-recursively, then an infinite co-recursive chain of
application of rule (4) would relate the silently diverging stream to any stream.

The three parameters to \texttt{bisimF} refine the way these rules can be used
to derive different relations. The boolean $b_L,~b_R$ flags enable or
disable rules (4) and (5) respectively. The $\betaclo$ parameter, of type
$\mathcal P(stream \times stream) \rightarrow \mathcal P (stream \times stream)$ is
slightly more subtle. When matching two external events by rule (3), one does
not have to relate the remaining of the streams with respect to just a
co-recursive call, but instead can first apply $\betaclo$ to it.

The practical use of the closure parameter will be delayed to
Section~\ref{sec:euttG} where it will be instrumental in deriving the necessary
reasoning principles. For now, we set the $\betaclo$ parameter to the identity
closure \texttt{id} in order to define the high level relations we are
interested in.
It is straightforward to check that $\mathtt{bisimF}~b_L~b_R~\betaclo$ is monotone
for any monotone $\betaclo$, in particular for $\mathtt{id}$.
We therefore can define the greatest fixed point $\mathtt{bisim}~b_L~b_R$ using \pacon.

We are now ready to derive concrete relations. First, if both asymmetric rules
are disabled, we have to exactly match all constructors: this corresponds to
strong bisimulation.

\begin{definition}[Strong bisimulation]
  \[s \bisim t \defeq \mathtt{bisim~\bfalse~\bfalse}~s~t \]
\end{definition}

At the opposite side, equivalence up-to-tau is defined by allowing both rules:
it is always fine to strip away finite amounts of $\tau$'s on either side:
\begin{definition}[Equivalence up-to-tau]
  \[s \eutt t \defeq \mathtt{bisim~\btrue~\btrue}~s~t \]

\end{definition}

Finally, a third relation is often useful. By allowing only one of the rules, we
get an asymmetric relation expressing that a stream is up-to-tau bisimilar to another,
but contains more $\tau$:
\begin{definition}[Over-approximation up-to-tau]
  \[s \euttge t \defeq \mathtt{bisim~\btrue~\bfalse}~s~t \]
\end{definition}

Notice the following subrelation inclusions: $\bisim\;\subseteq\;\euttge\;\subseteq\;\eutt$.

Unfortunately, the inductive-coinductive nature of weak bisimulation in particular
makes a property as elementary as transitivity already a challenge to prove. The
standard approach is to seek stronger reasoning principle by introducing
up-to techniques. We first consider reasoning up to transitive closure.

\subsubsection{Transitive Closure of the Bisimilarity Relations}

The native reasoning principle on bisimilarity only allows us to step through the
functor \texttt{bisimF}, forcing us systematically to nest an induction to
account for possible bounded stripping of $\tau$s, which often requires
a clever generalization of the statement for it to hold inductively. Reasoning
up-to transitive closure enables a new reasoning principle: when attempting to
prove that two streams $(s_1,s_2)$ belong to a relation $r$, it may be sound in
appropriate contexts to simply substitute $s_1$ or $s_2$ for other bisimilar
streams.

This intuition is formalized by introducing a family of transitive closures
parameterized by four booleans flags:

\begin{definition}[Transitive closure up to bisimilarity]
\[
\infer{(s_1,s_2)\in\mathtt{bisim\_trans\_clo}~b_L~b_R~b_L'~b_R'~r}{(s_1,s_1')\in\mathtt{bisim}~b_L~b_R\quad(s_1',s_2')\in r\quad (s_2,s_2')\in\mathtt{bisim}~b_L'~b_R'}
\]
\end{definition}

Each pair of flags defines the instances of \texttt{bisim} that are allowed to
be used to substitute for the left and right streams.
These closures are not all safe to use in arbitrary contexts. Indeed, by setting all
flags to $\btrue,$ we allow arbitrary rewriting \utt:

\begin{definition}[Undirected transitive closure]
  \[\transU \defeq \mathtt{bisim\_trans\_clo}~\btrue~\btrue~\btrue~\btrue\]
\end{definition}

Let us emphasize why such arbitrary, \emph{undirected}, up-to-tau rewriting
provided by $\transU$ is an unsound principle in general, which was first shown by~\citeN{Sangiorgi1992}.
Recall that a coinductive proof is in essence constructing a cycle by
being only allowed to invoke the coinduction hypothesis once below a guard.
In our case, $\transU$ could hence be misused to \emph{introduce}
a $\tau$ constructor that could then be used as a guard, allowing for unsound circular reasoning.
To illustrate the problem concretely, let us assume for a moment
that the precondition of the \textsc{Closure} principle from Figure~\ref{fig:gpaco-axiom} is available for $\transU$. The following proof would then be valid:
\begin{align*}
  &\Viss{0}\Ret\eutt \Viss{1}\Ret
  \iffv{\text{\textsc{Init}}}(\Viss{0}\Ret, \Viss{1}\Ret)\in \gpaconoclo{\euttF}{\emptyset}{\emptyset}\\
  \iffv{\text{\textsc{Acc}}}\quad& (\Viss{0}\Ret, \Viss{1}\Ret)\in\gpaconoclo{\euttF}{\emptyset}{\{(\Viss{0}\Ret, \Viss{1}\Ret)\}}\\
  \impliedbyv{\text{\textsc{Closure(\transU)}}}\quad& (\Tau\Viss{0}\Ret, \Tau\Viss{1}\Ret)\in\gpaconoclo{\euttF}{\emptyset}{\{(\Viss{0}\Ret, \Viss{1}\Ret)\}}\\
  \impliedbyv{Step}\quad& (\Viss{0}\Ret, \Viss{1}\Ret)\in\gpaconoclo{\euttF}{\{(\Viss{0}\Ret, \Viss{1}\Ret)\}}{\{(\Viss{0}\Ret, \Viss{1}\Ret)\}}\\
  \impliedbyv{Base}\quad& (\Viss{0}\Ret, \Viss{1}\Ret)\in\{(\Viss{0}\Ret, \Viss{1}\Ret)\}\qed
\end{align*}

This minimal example show-cases how this unrestricted up-to closure principle could
introduce \Tau constructors that would then be used as guards to wrongly justify the use
of the coinductive hypothesis. Thankfully, applying \textsc{Closure(\transU)} is prohibited.
Note however that had we justified the use of the coinductive hypothesis by a $\beta$ guard, the rewriting would have been harmless.

We will come back to $\transU$ in more detail by considering a
context-sensitive up-to technique in Section~\ref{sec:euttG}.
But let us focus for now on a better behaved instance:
\begin{definition}[Directed transitive closure]
  \[\transD \defeq \mathtt{bisim\_trans\_clo}~\btrue~\bfalse~\btrue~\bfalse\]
\end{definition}

The $\transD$ closure disables the second flag used in the setting of each
bisimulation considered. This means that a stream may be substituted by a
bisimilar one, only if the new one contains \emph{no more} $\tau$s than the previous
one. It is intuitively clear that this substitution is always sound since it
cannot introduce a guard.
Note that this is the up-to expansion technique presented by \citeN{Sangiorgi1992} to solve the problem of up-to weak bisimularity above.
This transitivity principle is in practice the most general one that we shall consider.
It will be the instance of the base closure that we will provide to \gpacon in the construction we introduce in Section~\ref{sec:euttG}.

This soundness and generality are expressed by proving that
$\transD$ provides a sound up-to reasoning principle with respect to $\eutt$.
This soundness holds in the sense that
$\transD$ satisfies the precondition from Lemma~\ref{lem:init} with respect to the functor
$\euttF \defeq \mathtt{bisimF~\btrue~\btrue}.$

Lemma~\ref{lem:init} allows us to move from a proof of a \pacon predicate, $\eutt$
being the one of concern, to a \gpacon counterpart setup with $\transD$ as the
base closure.
\begin{lemma}[Initialization for $\transD$ with respect to $\euttF$]
  For any monotone $\betaclo$ such that $\transD \circ \betaclo \subseteq \betaclo \circ \transD$,
  $\transD$ is weakly compatible for $\euttFg$.
\end{lemma}

We can at this stage already establish a certain number of facts about our instances of $\mathtt{bisim}.$
By picking in particular $\betaclo = \mathtt{id}$, the closure used in
the definition of $\euttF$, we can derive the following reasoning principle by applying \textsc{Closure*}.
\begin{theorem}[$\eutt$ is a congruence for $\euttge$]
  \[
  \infer{s\eutt t}{s' \euttge s \quad s'\eutt t' \quad t' \euttge t}
  \]
\end{theorem}
\noindent We then prove that $\mathtt{bisim}$ defines equivalence relations:
\begin{lemma}
  $\bisim$ and $\eutt$ are equivalence relations. $\euttge$ is reflexive and transitive.
\end{lemma}
\noindent And finally show that $\mathtt{bisim}~b_L~b_R$ is a congruence for each constructor of \euttF.

\subsubsection{Concat Closure}

Proving the monoidal laws and congruence rules relating \texttt{concat} to weak
bisimulation is greatly simplified by a second reasoning principle: the ability
to reason up-to prefix. When attempting to relate two streams defined as
concatenations, it should be possible to discharge their prefixes by proving they are
bisimilar. The following closure captures this reasoning principle:

\begin{definition}[Concat closure]
\[
\infer{(h_1\cat t_1,h_2\cat t_2)\in\concatC~r}
      {h_1\eutt h_2\quad(t_1,t_2)\in r}
      \]
\end{definition}

The soundness of the closure is embodied by showing that Lemma~\ref{lem:clo} can
be instantiated for $\concatC$ with respect to $\euttF$, with $\transD$ for
the base closure:
\begin{lemma}[Compatibility of $\concatC$ with respect to $\euttF$]
  \label{lem:concat-clo}
  For any $\betaclo$ monotone such that $\concatC \circ \betaclo \subseteq \betaclo \circ \concatC$ and $\mathtt{id}\subseteq \betaclo$,
    we have $\concatC \subseteq \gupaco{\euttFg}{\transD}{}$.
\end{lemma}

Lemma~\ref{lem:concat-clo} essentially states that all instances of
$\mathtt{bisim}$ are congruences for \texttt{concat} in the first argument. In
particular we can prove that $\bisim$ is a congruence for \texttt{concat}:

\begin{theorem}[$\bisim$ is a congruence for \texttt{concat}]
  \label{theo:bisim-concat}
  \[
  \infer{h_1\cat t_1\bisim h_2\cat t_2}{h_1 \bisim h_2 \quad t_1\bisim t_2}
  \]
\end{theorem}

With these tools in hand, we can prove the expected monoidal laws. In
particular, Theorem~\ref{theo:bisim-concat} greatly simplifies the proof of
associativity.
\begin{theorem}[(\texttt{stream},$\cat$) forms a monoid]
  \[\Ret\cat s\bisim s\qquad s\cat\Ret\bisim s\qquad (r\cat s)\cat t\bisim r\cat(s\cat t)\]
\end{theorem}
\section{An Equational Theory for Weak Bisimulations}
\label{sec:euttG}

Section~\ref{sec:bisim} introduced the \texttt{stream} data type and two
equivalence relations upon it: a strong bisimulation that constrains them to be
structurally identical, and a weak bisimulation that quotient them up-to finite
amount of internal steps. We have shown that two reasoning principles may be
proved sound when reasoning about weak bisimulations: up-to transitivity with
respect to addition of taus, $\transD$, and up-to concat closures, $\concatC$.

However, even with the support from \texttt{gpaco}, reasoning about streams
remains a technical challenge. In particular, we noticed that up-to transitivity
with respect to general equivalence \utt, $\transU$, is sound in contexts
guarded by a $\beta$, but not when guarded by a $\tau$.

In order to alleviate these difficulties, we abstract away from the low-level
use of \gpacon and define in this section a new \emph{context-sensitive}
weak bisimulation relation, $\mathtt{euttG}$. We prove that this relation satisfies a
rich equational theory, notably supporting context-sensitive up-to techniques,
and is sound with respect to weak bisimulation. By doing so, we hence
internalize much of the complexity inherent to coinductive reasoning over weak
bisimulation and provide an interface exposing the higher level reasoning principles specific to weak bisimulations of streams.

\subsection{A Context-Sensitive Weak Bisimulation}

We leverage the expressivity of \gpacon to define the
parameterized weak bisimulation $\euttGg.$
Before getting to its formal definition, we sketch the intuition it carries.
The relation takes four parameters, each of type $\mathcal P (\mathtt{stream} \times \mathtt{stream})$, which correspond respectively to information that has been
unlocked by a visible step or an internal step, or that remains guarded behind a
visible step or an internal step.

The key idea in distinguishing the kind of constructor that has released or
still guards the information is to allow for context-sensitive up-to techniques.
Indeed, an incremental coinductive proof can be thought as a game of exploration
whose goal is to close all paths explored by coming back to a previously
explored state. By substituting a stream for a weakly bisimilar one, we may
compromise all states reached by taking $\tau$ steps, but we remain certain
that a cycle is found if we get back to a state reached under a $\beta$ step.
As such, $\beta$ guards are stronger than $\tau$ guards when reasoning \utt.

The main tool we will use to enable more reasoning principles under
$\beta$ guards than $\tau$ guards is the $\betaclo$ argument introduced in the
definition of $\mathtt{bisim},$ Figure~\ref{fig:bisimgen}, and which has been
left unexploited through Section~\ref{sec:bisim}. Recall that this parameter is
a closure up-to which is applied to the co-recursive call under a $\beta$ constructor.
The closure we consider is defined as follows:

\begin{definition}[Closure for external events]
  \[
  \euttVC{\gs}~r \defeq \gupaco{\euttF~\mathtt{id}}{\transD}{\transU (r \cup \gs)}.
  \]
\end{definition}

The closure $\euttVC{\gs}$ is best understood right to left. At its core, it
simply extends the relation $r$ with the $\beta$ guarded knowledge
$\gs$. Since it will only be accessible under $\beta$ guards, it is also sound
to close this knowledge up to \emph{undirected} transitivity, $\transU$, to
allow for arbitrary rewriting by weak bisimilarity. Finally, by definition of
$\bisimF,$ using $\euttVC{\gs}$ in place of the $\betaclo$ argument permits its use
right as we strip off a pair of $\beta$ constructors. Specifically, if the goal
is of the form $\Vis{n}\cdot s\eutt\Vis{n}\cdot t,$ then $\euttVC{\gs}$ can be used to relate
$s$ and $t$. However, we sometimes want to delay the use of this
closure: say the goal is of the form $\Vis{n}\cdot p\cat s \eutt \Vis{n}\cdot p\cat t,$ we
need to first reason up-to concatenation and only then use $\euttVC{\gs}$ to relate
$s$ and $t$. Wrapping the whole closure into a call to \gpacon is a convenient
way to make this possible.

We now turn to the definition of $\euttGn$ itself:

\begin{definition}[Parameterized weak bisimulation]
  \[
  \euttGg \defeq
  \gpaco {\euttF~(\euttVC{\gs})}
         {\transD}
         {(\transU (\rs) \cup \rw))}
         {\gw}
  \]
\end{definition}

The definition of $\euttGn$ is a slightly intimidating instance of \gpacon. Let
us walk through each of its arguments. First, the base closure provided is $\transD$:
in any context, it is sound to work up to directed transitivity. Now since both
$\rs$ and $\rw$ are information that has been unlocked previously, their union
is provided as accessible, except that, as in the case of $\gs$ under $\euttVC{},$ the $\beta$
unlocked knowledge is additionally closed by $\transU$ -- undirected transitivity. The functor
whose greatest fixed point we take is naturally $\euttF;$ going under the
functor hence guarantees that we go either under a $\tau$ or a $\beta$ guard. We
therefore set $\gw$ to be \emph{always} unlocked under the functor, as expressed
by its position as last parameter of \gpacon. Finally, the additional
knowledge $\gs$ is ensured to be \emph{only} unlocked when the functor is
applied by going under $\beta$ guards by being provided as a parameter to
$\euttVC{}$ in the closure passed to $\mathtt{euttF}$.

Having motivated the definition of $\euttGn$ by the intuitive reasoning
principles it should satisfy, we formalize these principles in the following subsection.

\begin{figure*}
  \texttt{\bf{Soundness}} \\[0.1cm]
  \begin{centermath}
    \hfill
    \infer[\textsc{Init}]{s\eutt t}{(s,t)\in\euttG{\emptyset}{\emptyset}{\emptyset}{\emptyset}}
    \hfill
    \infer[\textsc{Final}]{(s,t)\in\euttGg}{s\eutt t}
    \hfill
  \end{centermath}
  \texttt{\bf{Knowledge~manipulation}} \\[0.1cm]
   \begin{centermath}
    \hfill
    \infer[\textsc{Base}]{(s,t)\in\euttGg}{(s,t)\in \rs\cup\rw}
    \hfill
    \infer[\textsc{Acc}]{x \subseteq \euttGg}{x \subseteq \euttG {\rs} {\rw} {(\gs \cup x)} {(\gw \cup x)}}
    \hfill
   \end{centermath}

   \texttt{\bf{Stream~processing}} \\[0.1cm]
   \begin{centermath}
     \infer[\textsc{Ret}]{(\Ret,\Ret)\in\euttGg}{}
     \hfill
     \infer[\tau\_\textsc{Step}]{(\tau\cdot s,\tau\cdot t)\in\euttGg}{(s,t)\in\euttG{\rs}{\gw}{\gs}{\gw}}
     \hfill
     \infer[\beta\_\textsc{Step}]{(\beta(n)\cdot s,\beta(n)\cdot t)\in\euttGg}{(s,t)\in\euttG{\gs}{\gs}{\gs}{\gs}}
   \end{centermath}

    \texttt{\bf{Up~to~reasoning}} \\[0.1cm]
    \begin{centermath}
      \infer[\textsc{TransD}]{(s,t)\in\euttGg}{(s,t)\in\transD(\euttGg)}
      \hfill
      \infer[\textsc{TransU}]{(s,t)\in\euttGg}{(s,t)\in\transU(\euttG{\rs}{\rs}{\gs}{\rs})}
      \hfill
      \infer[\textsc{ConcatC}]{(s,t)\in\euttGg}{(s,t)\in\concatC(\euttGg)}
   \end{centermath}
  \caption{Equational theory for parameterized equivalence up-to-tau. $\transD,$ $\transU$ and $\concatC$ are the closures for which up-to reasoning is possible: directed and undirected transitivity, and concatenation.}
  \label{fig:euttG-axiom}
\end{figure*}

\subsection{An Equational Theory for $\euttGn$}

The interface provided by our theory is summarized by the set of rules described in
Figure~\ref{fig:euttG-axiom}. They are split into four categories. The
\emph{soundness} rules relate equivalence \utt and
$\euttGn.$ The \emph{knowledge manipulation} rules provide the core coinductive
principles specialized to weak bisimulation. The \emph{stream processing} rules
give specialized principles to step under $\euttF$ constructors. Finally, we provide
support for three \emph{up-to} reasoning principles.
All rules maintain the following implicit invariant for \euttGn: $\rs \subseteq
\rw \subseteq \gw \subseteq \gs$.

\paragraph{Soundness}

The relation between $\euttGn$ and $\eutt$ is similar to the one between \pacon
and \gpacon: it is an intermediary construct one transits to in order to conduct
a proof.

The soundness of the overall approach is hence encapsulated into two rules.
First, the \textsc{Init} rule states that one can always move during a
proof of weak bisimulation into the $\euttGn$ realm by assuming no initial
knowledge.
\begin{theorem}[\textsc{Init}]
  \[
  (s,t)\in\euttG{\emptyset}{\emptyset}{\emptyset}{\emptyset} \implies s\eutt t
  \]
\end{theorem}

Using \textsc{Init}, we can hence start a $\euttGn$-based proof.
Conversely, since $\euttGn$ is purely an intermediary to conduct proofs about weak bisimulation,
\textsc{Final} is key to invoke any pre-established $\eutt$-equation:
for any state of accumulated knowledge, $\euttGn$ always contains $\eutt$.
\begin{theorem}[\textsc{Final}]
  \[
  s\eutt t \implies (s,t)\in\euttGg
  \]
\end{theorem}

\paragraph{Knowledge manipulation}

The $\euttGn$ relation shields the user from its internals as much as possible by
providing its own reasoning principles with respect to the four knowledge
arguments it carries. First, the \textsc{Base} case echoes its \gpacon
counterpart by giving access to all unlocked knowledge.
\begin{theorem}[\textsc{Base}]
  \[
  (s,t)\in \rs\cup\rw \implies(s,t)\in\euttGg
  \]
\end{theorem}

The accumulation theorem is once again key to make
parameterized coinductive reasoning possible. It states that in order to prove
that a set $x$ of pairs of streams belongs to $\euttGn$, one can
extend the guarded knowledge by assuming that $x$ is contained in
this knowledge:
\begin{theorem}[\textsc{Acc}]
  \[
  x \subseteq \euttGg \iff x \subseteq \euttG {\rs} {\rw} {(\gs \cup x)} {(\gw \cup x)}
  \]
\end{theorem}

\paragraph{Stream processing}

Three principles allow us to process each of the \texttt{stream} constructors.
Naturally, it is trivial to show that terminating streams can be matched.
\begin{theorem}[\textsc{Ret}]
  \[
  (\Ret,\Ret)\in \euttGg
  \]
\end{theorem}

Internal events can be consumed on each side, which grant access to the $\tau$ guarded knowledge.
\begin{theorem}[\textsc{$\tau$ step}]
  \[
  (t,s)\in \euttG{\rs}{\gw}{\gs}{\gw} \implies(\tau\cdot s,\tau\cdot t) \in \euttGg
  \]
\end{theorem}

Finally, visible steps propagate the guarded knowledge to all parameters.

\begin{theorem}[\textsc{$\beta$ step}]
  \begin{align*}
    &(t,s)\in \euttG{\gs}{\gs}{\gs}{\gs} \\
    &\implies(\beta(n)\cdot s,\beta(n)\cdot t) \in \euttGg
  \end{align*}
\end{theorem}

\paragraph{Up-to reasoning}

Finally, three up-to reasoning principles are supported. As developed in
Section~\ref{sec:bisim}, directed transitive closure and concatenation closure
are sound in all contexts. This gets reflected in the simplicity of rules
\textsc{transD} and \textsc{concatC}: one can simply make a call to the
corresponding closure at any time.

\begin{theorem}[Directed transitive closure]
  \[
  (s,t)\in\transD(\euttGg) \implies (s,t)\in \euttGg
  \]
\end{theorem}

\begin{theorem}[Concat closure]
  \[
  (s,t)\in\concatC(\euttGg) \implies (s,t)\in \euttGg
  \]
\end{theorem}

The third principle, undirected transitive closure, is more interesting.
We internalize the intuition that it is only sound while guarded by $\beta$ guards by overwriting all weakly available and guarded knowledge
by the strongly available one:

\begin{theorem}[Undirected transitive closure]
  \[
  (s,t)\in\transU(\euttG{\rs}{\rs}{\gs}{\rs}) \implies (s,t)\in \euttGg
  \]
\end{theorem}

We now illustrate a use of this interface.

\subsection{Practical Use of \euttGn}

Consider the following two streams:
  \[
  \begin{array}{@{}r@{\;}l@{\qquad}r@{\;}l@{\qquad}r@{\;}l@{\qquad}r@{\;}l@{}}
    s_0 &\eutt \Viss{0}\, s'_0 &
    s'_0 &\eutt r \cat s_1 &
    s_1 &\eutt \Viss{1}\, s'_1 &
    s'_1 &\eutt \Viss{2}\, s'_0
    \\
    t_0 &\eutt \Viss{0}\, t'_0 &
    t'_0 &\eutt r' \cat t_1 &
    t_1 &\eutt \Viss{1}\, t'_1 &
    t'_1 &\eutt \Viss{2}\, t'_0
  \end{array}
  \]
  This example differs from Figure~\ref{fig:ex-upto} in that each of the states
  are related to one another by \textit{weak} bisimilarity. To prove that $s_0
  \eutt t_0$ and $s_1 \eutt t_1$, the same proof as before using just \gpacon
  will not work, since we need to use $\transU$, a context-sensitive closure.
  However, the proof remains straightforward using \euttGn, assuming still that we know $r\eutt r'$,
  as depicted in Figure~\ref{fig:euttG-proof}.

\begin{figure}
\begin{align*}
  &X_0 \subseteq \gfp{\euttF}
  \impliedbyv{\text{\textsc{Init}}} X_0 \subseteq \euttG {\emptyset} {\emptyset} {\emptyset} {\emptyset}
  \\[-4pt]
  \impliedbyv{\text{\textsc{Acc}}}~ & X_0 \subseteq \euttG {\emptyset} {\emptyset} {X_0} {X_0}
  \\[-4pt]
  \impliedbyv{\text{\textsc{TransU}}}~ & \{ (0\, s_0', 0\, t_0'), (1\, s_1', 1\, t_1') \} \subseteq \euttG {\emptyset} {\emptyset} {X_0} {\emptyset}
  \\[-4pt]
  \impliedbyv{\text{$\beta$\_\textsc{Step}}}~ & X_1 \subseteq \euttG {X_0} {X_0} {X_0} {X_0}
  \\[-4pt]
  \impliedbyv{\text{\textsc{Acc}}}~ & X_1 \subseteq \euttG {X_0} {X_0} {(X_0 \cup X_1)} {(X_0 \cup X_1)}
  \\[-2pt]
  \mysmall{\textbf{lhs:}}~~~ & (s_0', t_0') \in \euttG {X_0} {X_0} {(X_0 \cup X_1)} {(X_0 \cup X_1)}
  \\[-4pt]
  \impliedbyv{\text{\textsc{TransU}}}~ & (r \cat s_1, r' \cat t_1) \in \euttG {X_0} {X_0} {(X_0 \cup X_1)} {X_0}
  \\[-4pt]
  \impliedbyv{\text{\textsc{ConcatC}}}~ & (s_1, t_1) \in \euttG {X_0} {X_0} {(X_0 \cup X_1)} {X_0}
  \\[-2pt]
  \impliedbyv{\text{\textsc{Base}}}~ & \specialcell{ (s_1, t_1) \in X_0 \hfill \qed}
  \\
  \mysmall{\textbf{rhs:}}~~~ & (s_1', t_1') \in \euttG {X_0} {X_0} {(X_0 \cup X_1)} {(X_0 \cup X_1)}
  \\[-4pt]
  \impliedbyv{\text{\textsc{TransU}}}~ & (2\, s_0', 2\, t_0') \in \euttG {X_0} {X_0} {(X_0 \cup X_1)} {X_0}
  \\[-4pt]
  \impliedbyv{\text{$\beta$\_\textsc{Step}}}~ & (s_0', t_0') \in \euttG {(X_0 \hspace{-2pt}\cup \hspace{-2pt}X_1)} {(X_0 \hspace{-2pt}\cup \hspace{-2pt}X_1)} {(X_0 \hspace{-2pt}\cup \hspace{-2pt}X_1)} {(X_0 \hspace{-2pt}\cup \hspace{-2pt}X_1)}
  \\[-4pt]
  \impliedbyv{\text{\textsc{Base}}}~ & \specialcell{ (s_0', t_0') \in X_0 \cup X_1 \hfill \qed}
\end{align*}
\caption{Practical use of \euttGn: a proof example}
\label{fig:euttG-proof}
\end{figure}

Notice in particular how \textsc{TransU} allows us to rewrite \utt equations, at
the cost each time of losing the knowledge locked behind a $\tau$ guard.

\subsection{Essential Need for the Base Closure}
\label{sec:need-for-bclo}

We show that the companion closure is inconsistent with the rules of $\euttGn$,
so that it cannot be used as a base closure.
To this end, for any definition of $\euttGn$ satisfying the rules in Figure~\ref{fig:euttG-axiom},
suppose that it is closed under the companion $\cpn_F$
for $F = \mathtt{bisimF}~b_L~b_R~\betaclo$ with arbitrary $b_L$, $b_R$, $\betaclo$:
\begin{eqnarray}
\cpn_F (\euttGg) \subseteq \euttGg \label{eqn:companion-clo}
\end{eqnarray}

\noindent Let $X = \{(\Viss{1} \app \Ret, \Viss{2} \app \Ret)\}$
and $Y = \{(\Viss{0} \app \Viss{1} \app \Ret, \Viss{0} \app \Viss{2} \app \Ret)\}$.
For $\top : \mathtt{stream} \times \mathtt{stream}$, we have:
\begin{eqnarray}
  \cpn_F(Y) = F(\top) \label{eqn:companion-prop} \\
  Y \subseteq \euttG {\emptyset} {\emptyset} {X} {\emptyset} \label{eqn:y-euttG-x}
\end{eqnarray}
The proof of (\ref{eqn:companion-prop}) is given in Appendix~\ref{sec:companion-prop}.
(\ref{eqn:y-euttG-x}) follows by applying $\beta\_\textsc{Step}$ then \textsc{Base}.

\begin{figure}
\begin{align*}
  &X \subseteq \gfp{F}
  \impliedbyv{\text{\textsc{Init}}} X \subseteq \euttG {\emptyset} {\emptyset} {\emptyset} {\emptyset}
  \\[-4pt]
  \impliedbyv{\text{\textsc{Acc}}}~ & X \subseteq \euttG {\emptyset} {\emptyset} {X} {X}
  \\[-4pt]
  \impliedbyv{\text{\textsc{TransU}}}~ & X \subseteq \transU (\euttG {\emptyset} {\emptyset} {X} {\emptyset})
  \\[-4pt]
  \impliedbyv{\text{by $(\ref{eqn:companion-clo})$}}~ & X \subseteq \transU (\cpn_F (\euttG {\emptyset} {\emptyset} {X} {\emptyset}))
  \\[-4pt]
  \impliedbyv{\text{by $(\ref{eqn:y-euttG-x})$}}~ & X \subseteq \transU (\cpn_F (Y))
  \\[-4pt]
  \iffv{\text{by $(\ref{eqn:companion-prop})$}}~ & X \subseteq \transU (F(\top))
  \\[-2pt]
  \impliedbyv{\text{}}~ & \specialcell{ X \subseteq \transU (F (X)) \text{ (since $(\Tau \Viss{1} \app \Ret, \Tau \Viss{2} \app \Ret) \in F(X)$)} \hfill \qed}
\end{align*}
\caption{A contradiction when the companion is used as the base closure}
\label{fig:cpn-contradiction}
\end{figure}

Then, as shown in Figure~\ref{fig:cpn-contradiction}, we can derive a contradiction, that $\Viss{1} \app \Ret \eutt \Viss{2} \app \Ret$.
The root of the issue is that the companion construction contains non-structural
``junk'' when provided a false assumption like $Y$ above. Where we would want
$\cpn_F(Y)$ to contain exactly the pairs of streams equivalent modulo $Y$, it
also ends up containing nonsensical pairs such as $(\Tau \Viss{1} \app \Ret,
\Tau \Viss{2} \app \Ret)$.



\section{Implementation in the Coq Proof Assistant and Large Scale Case-Study}
\label{sec:implementation}

We implemented \gpacon and its theory as described through
Section~\ref{sec:gpaco} in the Coq proof assistant. The formalization is built
as an extension of the \pacon library and available at \url{https://github.com/snu-sf/paco}.

Since the implementation builds directly on top of \pacon, it is
fully backward compatible: the new \gpacon reasoning principles are applicable
to any coinductive object defined via \pacon, with no change in the definitions.
As was the case with the original library, we provide high level
tactics mapping to each reasoning principle described in
Figure~\ref{fig:gpaco-axiom}.

\subsection{Large Scale Case-Study: Interaction Trees}
\label{sec:itrees}

For sake of exposition and self-containment, we have presented here a case-study
built on streams and their monoidal structure.
The motivation for the development of this technique however stemmed from a more
complex application: interaction trees~\cite{itrees} are a
coinductive structure similar to streams, but branching in the sense that the
visible events are followed by a continuation over the type of the emitted
event. Interaction trees can be equipped with a \texttt{bind} operation similar
to the \texttt{concat} operation, and proved to form a monad.

We have applied the techniques described in this paper to derive an axiomatic
interface to reason \utt about interaction trees. This layer of abstraction has
then been heavily used to reason about this structure, and proved instrumental
in alleviating the induced difficulty.

The corresponding formal development can be browsed at
\url{https://github.com/DeepSpec/InteractionTrees/}.
In particular, the equational theory is developed in the \texttt{/theories/Eq} directory.

\newcommand{\rel}[1]{\mathrel{\mathcal{#1}}}

\newcommand{\progress}[2]{#1\mathrel{\rightarrowtail}#2}
\newcommand{\dia}[3]{#1\mathrel{\twoheadrightarrow}#2,~#3}

\section{Discussion and Related Work}
\label{sec:related}

\paragraph{Paco and Companion}
We start by discussing how our contribution builds on existing works,
namely parameterized coinduction~(Paco)~\cite{paco} and the
companion~\cite{Pous16}, and how we improve on them.

As we reviewed in Section~\ref{sec:background}, Paco provides incremental
reasoning by the parameterized fixed point $\paco{f}{}$. It also provides up-to
reasoning by combining $f$ with its greatest respectful closure $\gres_f$
(\emph{i.e.}, using $\paco{f\,\circ\,\gres_f}{}$). \citeN{Pous16}
shows that the greatest compatible closure $\cpn_f$, called the \emph{companion},
coincides with $\gres_f$ and directly admits the incremental and up-to reasoning
principles of $\paco{f\,\circ\,\gres_f}{}$. Moreover, the companion admits second-order
reasoning, which provides incremental and up-to principles for reasoning about
$\mathit{clo} \sqsubseteq \cpn_f$.

In our work, we generalize the constructions in two directions.
First, we use two parameters to track both the unlocked and guarded knowledge.
As briefly discussed in Section~\ref{sec:up-to}, the companion construction with two parameters
$r$ and $g$ can be given by $\cpn_f(r \sqcup f(\cpn_f(r \sqcup g))$.
Second, we parameterize the upper-bound of closures
instead of using the greatest compatible/respectful closure.
The need for such parameterization was shown in Section~\ref{sec:need-for-bclo}.

\paragraph{Distinguishing Internal and Visible Steps}
\cite[Exercise 2.4.64]{SangiorgiWalker} and \cite{Pous07} present
up-to techniques allowing different up-to closures for internal and
visible steps. Among them, \cite{Pous07} gives a more formal framework,
where two notions of monotonicity (in a more recent terminology, respectfulness)
are defined. If a relation $R$ is $\Tau$-simulated (\emph{i.e.}, for internal steps)
up-to a monotonic closure and \texttt{v}-simulated (\emph{i.e.,} for visible steps)
up-to a weakly monotonic closure, then $R$ is contained in the weak (bi)similarity.
Notably, up-to weak bisimulation is only weakly monotonic.

Similarly, our work also presents an equational theory for weak
bisimulation where internal and visible steps admit different up-to
closures. The main challenge we are addressing is to combine such
up-to closures with incremental reasoning using four different kinds
of knowledge: unlocked/guarded knowledge for internal/visible steps.

\citeN{ABLP16} have developed a general framework to reason about
notions of weak steps vs. strong steps (passive vs. active in their terminology)
when establishing a bisimulation.
Simulations can generally be phrased in term of a relation $\rel R$ that
progresses to itself: \(\progress{\rel{R}}{\rel{R}}\). Under this formulation, an
up-to technique is a function $f$ on relations such that when \(\progress{\rel
  R}{f(\rel R)}\), then $\rel R$ is included in the bisimilarity relation.
In order to account for a distinction of the stepping relation between a passive part
and an active part, they introduce the notion of diacritical progress:
$\dia{\rel R}{\rel Q}{\rel S}$ expresses that $\rel R$ progresses toward $\rel
Q$ in the passive case, toward $\rel S$ in the active case.
With this tool, an up-to technique in the usual sense (called strong) is a function $f$ such
that $\dia{\rel R}{f(\rel R)}{f(\rel R)}$ implies that $\rel R$ is in the
bisimilarity relation.
This definition also extends to functions $f$ such that
$\dia{\rel R}{\rel R}{f(\rel R)}$ implies the same.
These up-to techniques make explicit the fact that up-to reasoning is only enabled when performing active steps.
In ~\cite{ABLP16}, they develop sufficient conditions for using strong and regular up-to techniques
in terms of the notions of evolution and compatibility of functions, adapted to the diacritical setting.
\cite{diacritical} goes further by generalizing this view to the lattice-theoretic
setting. This generalization allows them to introduce a notion of diacritical
companion defined as the greatest diacritically compatible function, extending
on both their and Pous' work.

This approach, whose contribution is orthogonal to that of this paper, we conjecture could be defined in \gpacon.
The development of $\euttGn$, and of the soundness of the $\texttt{transU}$ rule
in particular, might then fit nicely into this framework, potentially benefiting from this more principled approach in being easier to define.
Investigating this conjecture formally would be an interesting approach for future work.

\paragraph{Other Related Works}

In \cite{Pous16}, Pous introduced the companion of a function $f$ by
characterizing it as the greatest compatible function for $f$. \citeN{PW16} give a more explicit,
ordinal-based construction of the companion in classical set theory.
Analogously, it turns out that the companion can be obtained in constructive
type theory with an inductive tower construction as studied by Schäfer et
al.~\cite{SSD15,SchaferThesis}.

\cite{danielsson2017} presents a class of up-to techniques using size-preserving functions, which use sized types to prove the soundness of the techniques.
This class of techniques is shown to be related to Pous' companion, but does not include some useful up-to techniques.
Namely, Danielsson shows that techniques related to transitivity, such as those discussed in this paper, do not easily fit into the framework of size-preserving functions.

We have chosen to build our approach on top of \pacon, but other incremental coinductive
techniques exist: incremental pattern-based coinduction~\cite{Popescu10},
circular coinduction~\cite{HMS05}, parametric coinduction~\cite{Moss01}.
We refer to Hur et al.'s related work~\cite{paco} for a thorough comparison.

Finally, we introduced through this paper the use of three up-to techniques
relevant to our domain of application.
Numerous others can be found in Pous~\cite{Pous16}, both derived from the companion and as part of the related work.






\appendix
\section{Appendix}
\subsection{A Property about the Companion}
\label{sec:companion-prop}

Let $X = \{(\Viss{1} \app \Ret, \Viss{2} \app \Ret)\}$
and $Y = \{(\Viss{0} \app \Viss{1} \app \Ret, \Viss{0} \app \Viss{2} \app \Ret)\}$.
We prove that $\cpn_F(Y) = F(\top)$
for $F = \mathtt{bisimF}~b_L~b_R~\betaclo$ with arbitrary $b_L$, $b_R$, $\betaclo$.

We first define a function $\mathit{clo}$ as follows:
\[
\mathit{clo}(r) =
\left\{
\begin{array}{rl}
  \top & \text{if $X \subseteq r$} \\
  F(\top) & \text{else if $Y \subseteq r$} \\
  \emptyset &  \text{otherwise}
\end{array}
\right.
\]
Then $\mathit{clo}$ is trivially monotone and compatible as follows.
For any $r$, we show $\mathit{clo} (F(r)) \subseteq F (\mathit{clo}(r))$ by case analysis on $r$.
First, when $X \subseteq r$, we have $\mathit{clo}(r) = \top$.
We also have $Y \subseteq F(X) \subseteq F(r)$ and $X \not\subseteq F(r)$ by definition of $F$.
Therefore, we have $\mathit{clo}(F(r)) = F(\top) = F(\mathit{clo}(r))$.
Second, when $X \not\subseteq r$,
we have $X \not\subseteq F(r)$ and $Y \not\subseteq F(r)$ by definition of $F$.
Therefore, we have $\mathit{clo}(F(r)) = \emptyset \subseteq F(\mathit{clo}(r))$.

Now, we have the following inequality:
\[
\begin{array}{@{}r@{~}c@{~}l@{\quad}l@{}}
  F(\top) &=& \mathit{clo}(Y)  & \text{(by definition of $\mathit{clo}$)} \\[1mm]
  &\subseteq& \cpn_F(Y) & \text{($\cpn_F$ includes every compatible func.)} \\[1mm]
  &\subseteq& \cpn_F(F(X)) & \text{(by definition of $F$)} \\[1mm]
  &\subseteq& F(\cpn_F(X)) & \text{($\cpn_F$ itself is compatible)} \\[1mm]
  &\subseteq& F(\top)
\end{array}
\]
Therefore, we have $\cpn_F(Y) = F(\top)$.



\begin{acks}
This work was funded by the National Science
Foundation's Expedition in Computing \emph{The Science of Deep Specification}
under award 1521539 (Weirich, Zdancewic, Pierce)
with additional support by the ONR grant {\em REVOLVER} award N00014-17-1-2930,
and by the Basic Science Research Program through the National Research Foundation of Korea (NRF) funded by the Ministry of Science and ICT (2017R1A2B2007512).
We are grateful to all the members of the DeepSpec project for
their collaboration and feedback, and
we greatly appreciate the reviewers' comments and suggestions.

\end{acks}

\balance
\bibliography{ref}

\end{document}